\documentclass[aps,prl,twocolumn,superscriptaddress]{revtex4-2}

\usepackage{
	graphicx,		% For image modifications and the figure enviroment
	amsmath,		% For the AMS math styles
	amssymb,		% The extended AMS math symbol list
	tikz,			% Difficult drawing of awesome vector plots
	listings,		% For listing pieces of code in a nice and neat way
	hyperref,		% Makes links, references, the Table of Contents, etc. clickable.
	bbold,			% Allows the use of a special kind of bold notation in math mode
	subfigure
}

\begin{document}

%\preprint{}

\title{A comparison of cluster algorithms for the bond-diluted Ising model}


\author{Arnold H. Kole}
%\thanks{}
\affiliation{Department of Information and Computing Sciences, Utrecht University, Princetonplein 5, 3584 CC Utrecht, The Netherlands}
\affiliation{Institute for Theoretical Physics and Center for Extreme Matter and Emergent Phenomena, Utrecht University, Princetonplein 5, 3584 CC Utrecht, The Netherlands}

\author{Gerard T. Barkema}
\affiliation{Department of Information and Computing Sciences, Utrecht University, Princetonplein 5, 3584 CC Utrecht, The Netherlands}

\author{Lars Fritz}
\affiliation{Institute for Theoretical Physics and Center for Extreme Matter and Emergent Phenomena, Utrecht University, Princetonplein 5, 3584 CC Utrecht, The Netherlands}


\date{\today}

\begin{abstract}
Monte Carlo cluster algorithms are popular for their efficiency in
studying the Ising model near its critical temperature. We might
expect that this efficiency extends to the bond-diluted Ising model.
We show, however, that this is not always the case by
comparing how the correlation times $\tau_w$ and $\tau_{\rm sw}$ of
the Wolff and Swendsen-Wang cluster algorithms scale as a function of
the system size $L$ when applied to the two-dimensional bond-diluted
Ising model. We demonstrate that the Wolff algorithm suffers from
a much longer correlation time than in the pure Ising model, caused
by isolated (groups of) spins which are infrequently visited by
the algorithm. With a simple argument we prove that these cause the correlation
time $\tau_w$ to be bounded from below by
$L^{z_w}$ with a dynamical exponent $z_w=\gamma / \nu\approx 1.75$ for a bond concentration
$p < 1$. Furthermore, we numerically show that this lower bound is actually
taken for several values of $p$ in the range $0.5 < p < 1$. Moreover,
we show that the Swendsen-Wang algorithm does not suffer from the same problem.
Consequently, it has a much shorter correlation time,
shorter than in the pure Ising model even. Numerically at $p = 0.6$,
we find that its dynamical exponent is $z_{\rm sw} = 0.09(4)$.
\end{abstract}

%\keywords{}

\maketitle

\section{Introduction\label{sec:intro}}

% Context/Importance of bond-diluted Ising model
The Ising model is one of the most popular models in statistical physics:
its simplicity makes it easy to study while it is complex enough
that many interesting physical phenomena can be studied with it, such as
phase transitions and criticality \cite{brush1967history}. Since its
inception, numerous variants of the Ising model have been proposed to study
different phenomena. An important class of such variants are the Ising
models with impurities. These are used to investigate how the presence
of impurities, which occur frequently in nature, affects the properties
of a system. Common ways to model impurities in the Ising model is by
randomly removing spins (site-dilution \cite{hasenbusch2007universality,
ballesteros1997ising, ivaneyko2005criticality}), bonds (bond-dilution
\cite{Zhong_2020, Hasenbusch_2008, PhysRevB.18.2387, HADJIAGAPIOU20111279,
berche2004bond}) or alternatively by randomly modifying the strength of the interactions
in some other way \cite{Hasenbusch_2008, wolff1983phase}. In this paper we
focus on the variant with bond-dilution.

% What has already been discovered about the bond-diluted Ising model
The introduction of bond-dilution to the Ising model changes its properties
significantly. For example, it has been shown that the critical
temperature that separates the ferromagnetic and paramagnetic phases of
the Ising model changes depending on the extent of the bond-dilution
\cite{HADJIAGAPIOU20111279}. This even introduces a new type of phase
transition because the critical temperature drops to zero at a certain
bond concentration creating two phases (zero and non-zero critical
temperature) separated by what is referred to as the percolation
threshold \cite{jain1995anomalously}. In addition, it appears that the
presence of impurities also alters the universality class of the model
\cite{hasenbusch2007universality}.

% Importance of cluster algorithms for bond-diluted Ising model + what
% research has already been done on this and what we did
A common approach to study the Ising model is the use of Monte Carlo
methods. The choice of the algorithm does not change any of the equilibrium properties:
all algorithms sample the same (Boltzmann) distribution. However,
the dynamics of different algorithms can vary strongly leading to pronounced
differences in their efficiency for studying a certain model. In the
pure Ising model, cluster algorithms such as the Wolff and Swendsen-Wang algorithms
have proven themselves to be much more effective at criticality than
single spin-flip algorithms like Metropolis \cite{BarkemaBook}. This
difference is expected to be even more pronounced in the bond-diluted
Ising model since it has been recently shown that single spin-flip algorithms suffer
from a diverging correlation time when the percolation threshold is approached
\cite{Zhong_2020}. The dynamics of cluster algorithms for
the bond-diluted Ising model remains poorly studied and so it is still
unclear whether they actually are more effective. Some studies have proposed
that the efficiency of these cluster algorithms carries over to the
bond-diluted Ising model and that correlation times actually decrease
when site- or bond-dilution is introduced \cite{hennecke1993critical,
ivaneyko2005criticality}. We present a quantitative analysis of the
dynamics of the Wolff and Swendsen-Wang algorithms to show that this is
in fact not the case for the Wolff algorithm. We will demonstrate that
the Wolff algorithm suffers from much longer correlation times than in
the pure model, caused by isolated (groups of) spins, a fact which has
previously been hinted at by Ballesteros et al,
who showed that depending on the degree of bond dilution there are different
regions, characterised by the size of the groups of isolated spins, where
certain Monte Carlo updates are more efficient at thermalising the system \cite{ballesteros1998critical}.
We expand upon their work by proving a lower bound on the dynamical exponent
of the Wolff algorithm and numerically showing that this lower bound is actually
taken for several values of the dilution.

% Organisation of paper (how to refer to sections when there are no numbers)
This paper is organised as follows. We first define the bond-diluted
Ising model, the cluster algorithms, and the observables that we
use. Next, we present our results and discuss what they teach us about
the correlation times of the Wolff and Swendsen-Wang algorithms. In the
final section we summarise our main findings and conclude.

\section{Model and Methods\label{sec:modelmethods}}

\subsection{Model}

% Define the model
In this paper we study the bond-diluted Ising model in two dimensions on
a square lattice of size $L\,\times\,L$. This model is a variant of the
regular Ising model with nearest-neighbour interactions and is obtained
by randomly removing a fraction $1 - p$ of the bonds (i.e. interactions
between two neighbours) from the lattice, where $p$ is called the bond
concentration. Defined this way, $p$ is the probability
that there is a bond between two neighbours. With this choice, $p = 1$
corresponds to the regular Ising model and $p = 0$ to a collection of
isolated spins (no interactions). We define the model with the
Hamiltonian

\begin{equation}\label{eq:hamiltonian}
	\mathcal{H} = -J \sum_{\langle ij \rangle} c_{ij}(p) s_i s_j
\end{equation}

% Stuk over frozen-in bond-dilution duidelijk genoeg?
\noindent
where the sum runs over all pairs of nearest-neighbour sites, $s_i = \pm 1$
is the spin on site $i$ and $c_{ij}(p)$ is a constant that follows a
Bernoulli distribution with probability $p$, i.e. it has value $1$ with
probability $p$ and value $0$ with probability $1 - p$. We refer
to a realization of the $c_{ij}$'s for all nearest-neighbour pairs
as a configuration of the model. The bond-dilution is frozen in for a
particular configuration. In other words, the values of the $c_{ij}$'s
are fixed for a specific configuration. All through the manuscript, energy is measured in units of $J$.

\subsection{Algorithms}

% Describe both (Wolff and Swendsen-Wang) algorithms
We use the bond-diluted Ising model to study the behaviour, and in
particular the dynamics, of two cluster Monte Carlo algorithms. The first
of these is the Wolff algorithm \cite{PhysRevLett.62.361}. The basic idea behind this algorithm
is to grow a cluster of spins and flip all the spins in this cluster
simultaneously with probability 1. To grow a cluster we perform the
following steps \cite{BarkemaBook},

\begin{enumerate}
	\item choose a spin at random from the lattice,
	\item consider each of its neighbours. If the spins are aligned, add the neighbour to the cluster with probability $1 - e^{-2\beta J}$ with $\beta = \frac{1}{k_B T}$ and $J$ the coupling constant from the Hamiltonian,
	\item for each of the neighbours added in step 2 also consider all their neighbours to be added to the cluster and repeat this until no more neighbours exist that have not yet been considered.
\end{enumerate}

\noindent
It can be shown that by growing the cluster in this way we satisfy both ergodicity and detailed balance \cite{BarkemaBook}. It is important to note that in the bond-diluted Ising model two spins are only considered to be neighbours if there is a bond between them.

The second algorithm under consideration is the Swendsen-Wang
algorithm \cite{PhysRevLett.58.86}. Similar to the Wolff algorithm, clusters of spins are grown
according to the aforementioned procedure. It differs, however, in
the fact that we do not just grow a single cluster, but cover the
entire lattice with clusters and flip each of these with probability
$\frac{1}{2}$ in a single step \cite{BarkemaBook}. Since clusters are
grown in the same way as in the Wolff algorithm, showing that the Swendsen-Wang algorithm
satisfies ergodicity and detailed balance proceeds analogously \cite{BarkemaBook}.

\subsection{Observables\label{sec:observables}}

% Shortly mention the common ones (magnetisation, energy)
During our simulations we keep track of several quantities. This includes
the energy of a state, which follows directly from the
definition of the model and requires no further explanation. Additionally,
we measure a quantity which we will refer to as the spin
age and which we define as follows.

% Explain the spin age
To extract more information about the dynamics of the Wolff algorithm
from our simulations, we label each site in the lattice with a spin age
$a_i$, which we define to be the time since site $i$ was last visited
(i.e. was part of a Wolff cluster) measured in the number of Wolff
cluster moves. In other words, when a site is visited, its age is set
to $0$ and each subsequent Wolff cluster move where the site is not
visited, the age is incremented by $1$. Once the system is thermalised,
both with respect to its configuration of spins and the distribution
of ages, we count how often a certain age occurs at various steps
in the simulation, to produce a histogram showing the distribution of
ages in equilibrium. To be specific, at certain steps in the simulation
(between moves) we measure for each age $a$ how many spins in the lattice
are labelled with that age at that step and we call this number the age
frequency $f_L (a)$.

\section{Results and Discussion\label{sec:resultsdiscussion}}
\subsection{The Wolff algorithm}

We first discuss the behaviour of the Wolff algorithm applied to the
bond-diluted Ising model. We will start with a simple argument to show
that there must be a lower limit on the correlation time. Then we discuss
the results from our numerical analysis to show that this lower bound
is also taken for several values of the bond concentration $p$.
But before going into the simple argument, let us first introduce some
notation and two different time scales that we have used.

For all the results we measure time in cluster moves of the
algorithm used, because we found this the most intuitive timescale for
understanding the results. However, when evaluating the performance of an
algorithm, we prefer to measure time such that it scales with required
CPU time. Since the CPU time per single Wolff cluster move can vary
significantly, we require a second timescale for the Wolff algorithm.
A good candidate is to measure time such that $t = 1$ corresponds to the
situation where on average as many spins are flipped as there are in the lattice.
The relation between the time $t$
and our previous time, which we denote by $t_{\mathrm{steps}}$ for
Wolff, is given by $t = t_{\mathrm{steps}} \frac{\langle n \rangle}{L^2}$,
where $\langle n \rangle$ is the average size of a Wolff cluster
\cite{BarkemaBook}. It can also be shown that $\langle n \rangle$
scales as $L^{\gamma / \nu}$ at the critical temperature such that $L^{\gamma / \nu - 2}$ acts as a conversion factor when required
\cite{BarkemaBook}. By construction, the same number of spins, namely all the
spins in the lattice, are visited by the Swendsen-Wang algorithm in each
cluster move, so $t_{\mathrm{steps}}$ already scales with CPU time for
Swendsen-Wang and there is no additional timescale meaning we
use $t$ to denote the time measured in Swendsen-Wang cluster moves.

Now let us turn to the simple argument.
We argue that the correlation time  $\tau_{\mathrm{steps}, w}$ for $p < 1$ is bounded from
below by $L^2$, i.e. $\tau_{\mathrm{steps}, w} = \Omega(L^2)$. To see this,
note that for any $p < 1$ there will always exist at least one isolated spin
in the lattice for a sufficiently large system size (i.e. for a sufficiently
large system the expectation value for the number of isolated spins will be at least 1).
With an isolated spin we mean spins which have all their bonds to the rest
of the lattice removed. Such spins would only be flipped by the Wolff algorithm if
they are chosen as the seed spin. And since each spin is equally likely
to be picked and there are $L^2$ spins, the
correlation stored in these spins, however small it might be, will also
survive for $\Omega(L^2)$ cluster moves. Therefore, we can conclude
that the correlation time $\tau_{\mathrm{steps}, w}$ is bounded from below
by $L^2$.

To study the behaviour numerically, we ran simulations with
the Wolff algorithm for various system sizes with $p = 0.6$ at $(\beta
J)^{-1} = 0.940$ where $\beta = \frac{1}{k_B T}$ and $J$ the coupling
constant. We chose this value for $p$ because the effects of bond-dilution
become more pronounced when the bond fraction $p$ is significantly
below $1$. The temperature was chosen to be in the vicinity
of the critical temperature as determined with the Binder cumulant. The
value we found is also in good agreement with the critical temperature
found in other papers, see for example \cite{Zhong_2020}. Unless otherwise
mentioned, we used $100,000$ different realizations of the bond dilution in each
simulation. 

\begin{figure}
	\includegraphics[width=\columnwidth]{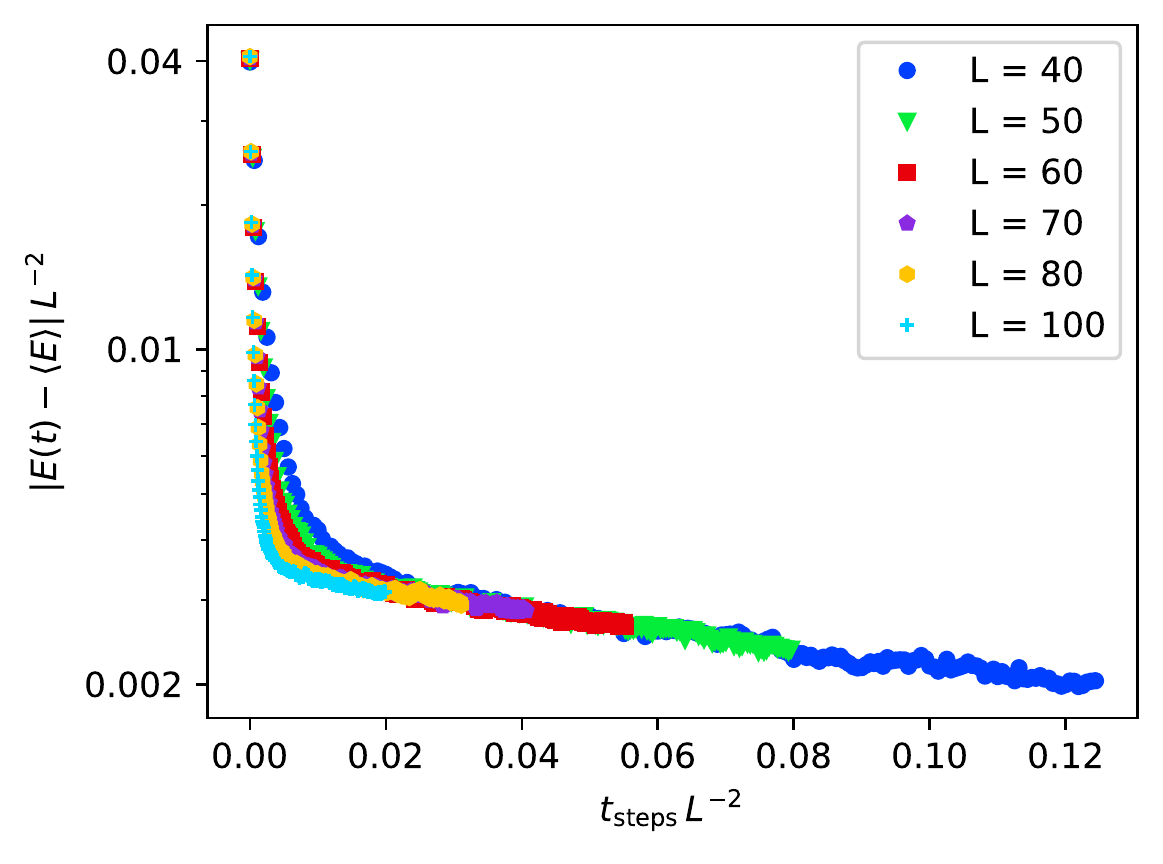}
	\caption{\label{fig:wolff_therm} Convergence of the energy
	$E(t)$ to the thermal equilibrium $\langle E \rangle$ during
	thermalisation with the Wolff algorithm for different system
	sizes $L$ with $p = 0.6$ at $(\beta J)^{-1} = 0.940$ where
	$\beta = \frac{1}{k_B T}$ and $J$ the coupling constant. For
	$t_{\mathrm{steps}} = 0$ the system starts in a state with
	all spins pointing up. Both the vertical and horizontal
	axes were scaled with $L^2$. Note the collapse of the right
	tails of the curves, suggesting that the correlation time
	$\tau_{\mathrm{steps}, w}  \sim L^2$.}
\end{figure}

Figure \ref{fig:wolff_therm} shows the evolution of the energy of
the system towards its thermal equilibrium value as a function of
Wolff cluster moves. For $L = 40$ we ran for 400 cluster moves per
configuration, for $L = 100$ we ran for 300 cluster moves and in
between we tuned the number of cluster moves to roughly keep the CPU time
used per simulation constant. At $t_{\mathrm{steps}} = 0$, the system
starts in the configuration with all spins pointing up ($s_i = 1$ for all
$i$). Notice how the curve seems to transition from a fast decay for small
$t_{\mathrm{steps}}$ to a slower decay at large $t_{\mathrm{steps}}$. When
the vertical and horizontal axes are scaled with $L^2$ the tails of
the curves, the regions of slower decay, collapse. Since these tails
are the limiting factor in convergence of the energy to its equilibrium
this suggests that the correlation time $\tau_{\mathrm{steps}, w}$
scales as $L^2$ such that $\tau_w$ scales as $L^{z_w}$ with $z_w={\gamma /
\nu}$. Numerically, it is reported that $\gamma / \nu$ is independent of
$p$ for $p \geq 0.6$, and actually indistinguishable from  $\gamma / \nu
= 1.75$ as in the regular Ising model \cite{HADJIAGAPIOU20111279}. Note
that, while the equilibrium exponents are numerically indistinguishable,
the dynamic exponent is very different: in the regular two-dimensional (2D) Ising model
the dynamic exponent is reported as $z_w = 0.25(1)$ \cite{BarkemaBook}.

% Plot to illustrate our suspicions of a problem with Wolff (convergence of <E> during thermalisation to equilibrium value maybe?)

\begin{figure*}
	\subfigure[$p < 1$]{\includegraphics[width=\columnwidth]{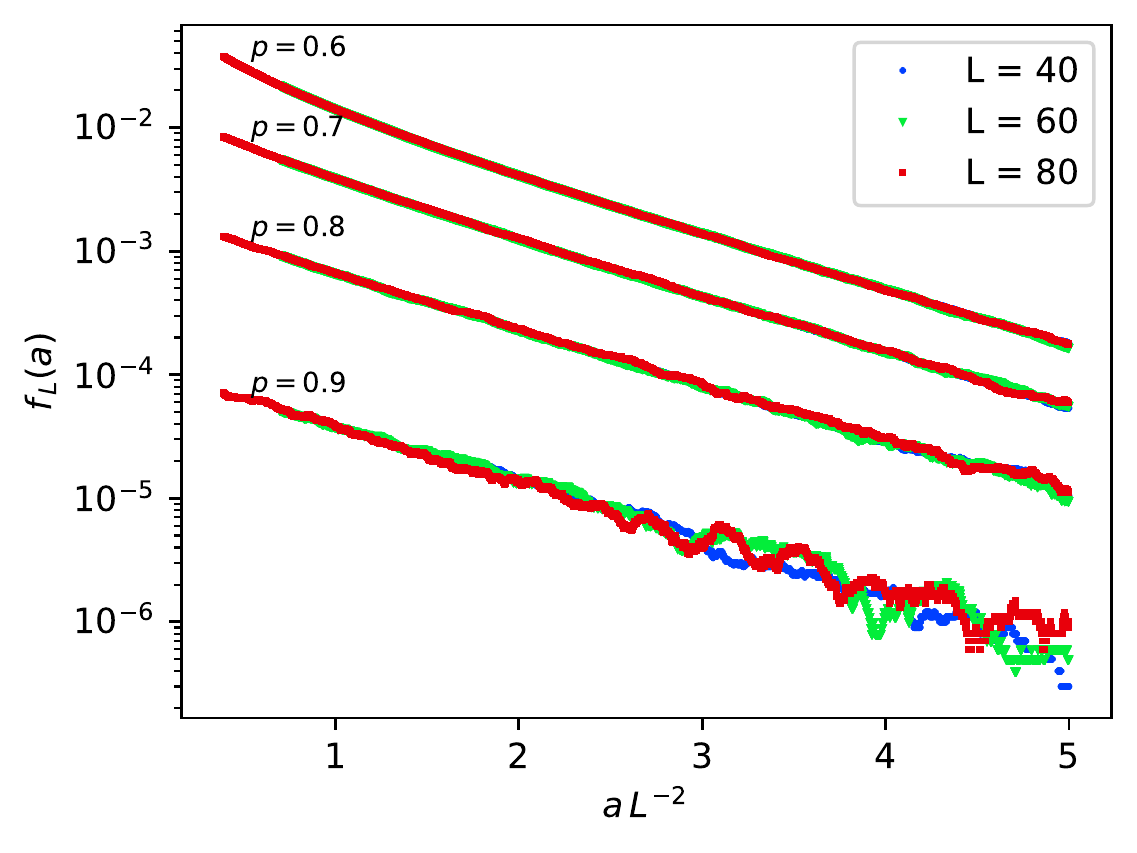}}\quad
	\subfigure[$p = 1$]{\includegraphics[width=\columnwidth]{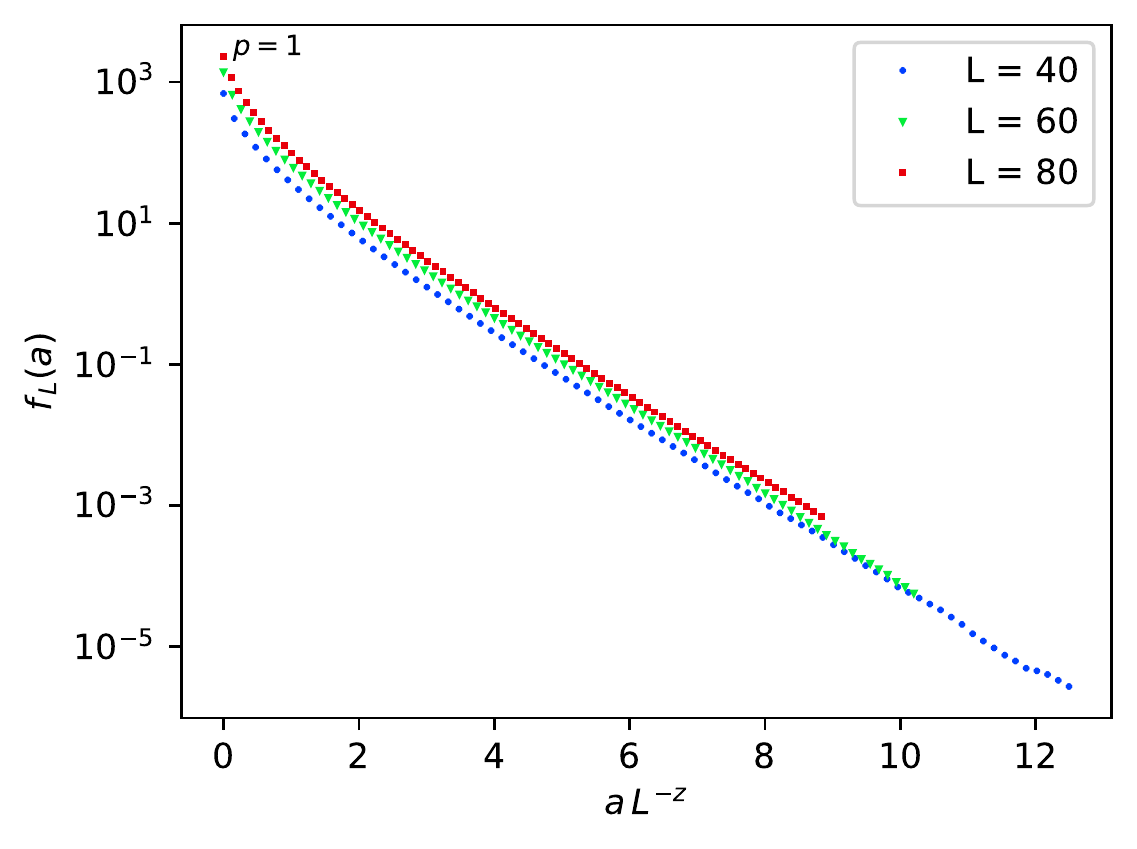}}
	\caption{\label{fig:wolff_histogram} Distribution of spin ages
		$a$ during a simulation with the Wolff algorithm at equilibrium. In figure (a) we see the data for $p = 0.6$ at $(\beta J)^{-1} = 0.940$, $p = 0.7$ at $(\beta J)^{-1} = 1.310$, $p = 0.8$ at $(\beta J)^{-1} = 1.648$ and $p = 0.9$ at $(\beta J)^{-1} = 1.964$. In figure (b) we see the data for $p = 1$ at $(\beta J)^{-1} = 2.27$. Here $\beta = \frac{1}{k_B T}$
		and $J$ the coupling constant. The spin age is defined as
		the time since the site was last visited, measured in Wolff
		cluster moves. Note the different scaling of the horizontal axis for (a) and (b). The horizontal axis in (a) was scaled with $L^2$, while in (b) it was scaled with $L^{z_{\mathrm{steps}, w}}$ where $z_{\mathrm{steps}, w} = 0.50$ was chosen to correspond with the $z_w = 0.25(1)$ for the regular 2D Ising model \cite{BarkemaBook}. The
		collapse of the curves in (a) again suggests that the correlation time
		$\tau_{\mathrm{steps}, w}$ of the Wolff algorithm scales as $L^2$,
		in agreement with figure \ref{fig:wolff_therm}. Scaling the horizontal axis in (b) with the dynamical exponent
		for the regular 2D Ising model from the literature also leads to a reasonable collapse, as we would expect.}
\end{figure*}

To verify our argument that isolated (groups of) spins exist that
are not touched by the algorithm for a long time, we computed a histogram of
the distribution of the spin ages throughout a simulation with the Wolff
algorithm in the manner described in the Model and Methods section. For
these simulations we used $10^4$ realizations of the bond-dilution. To initialise
the system we first thermalise with 50 Swendsen-Wang moves, starting from
a state with all spins pointing up. We also first run the simulation for
$5L^2$ Wolff cluster moves to make sure that spins can actually reach all
the ages that we report in the histogram. Finally, we measure the age for
an additional $1000$ consecutive Wolff steps. We did the simulations for
both $p = 0.6$ at $(\beta J)^{-1} = 0.940$ as before as well as for $p = 0.7$
at $(\beta J)^{-1} = 1.310$, $p = 0.8$ at $(\beta J)^{-1} = 1.648$ and $p = 0.9$ at $(\beta J)^{-1} = 1.964$.
For completeness, we also did the simulations at $p = 1$ at $(\beta J)^{-1} = 2.27$. 
We found these temperatures to be in the vicinity of the
critical temperature at their respective bond fractions $p$, again in agreement with
the critical temperature found in other papers \cite{Zhong_2020}.
The results are shown in figure \ref{fig:wolff_histogram}.

%\begin{figure}
%	\includegraphics[width=\columnwidth]{wolff_histogram.pdf}
%	\caption{\label{fig:wolff_histogram} Distribution of spin ages
%	$a$ during a simulation with the Wolff algorithm at equilibrium
%	with $p = 0.6$ at $(\beta J)^{-1} = 0.940$ and with $p = 0.7$
%	at $(\beta J)^{-1} = 1.310$ where $\beta = \frac{1}{k_B T}$
%	and $J$ the coupling constant. The spin age is defined as
%	the time since the site was last visited, measured in Wolff
%	cluster moves. The horizontal axis was scaled with $L^2$. The
%	collapse of the curves again suggests that the correlation time
%	$\tau_{\mathrm{steps}, w}$ of the Wolff algorithm scales as $L^2$,
%	in agreement with figure \ref{fig:wolff_therm}.}
%\end{figure}

% Mention intuition of how this leads to a tau_steps correlation of L^2 here.
The figure clearly shows that some spins 
survive for a very long time. Also note the strikingly good collapse of
the curves in \ref{fig:wolff_histogram}a when we scale the horizontal axis with $L^2$, for $p =
0.6$, $p = 0.7$, $p = 0.8$ and $p = 0.9$. This supports our earlier finding that $\tau_w$ scales
as $L^{z_w}$ with $z_w=\gamma / \nu \approx 1.75$ for these values of $p$. In contrast,
the histogram drops to zero very quickly for $p = 1$ and we need a different scaling to get
a reasonable collapse. This seems to suggest that the effect of long surviving spins only shows up for $p < 1$.

%It also seems to suggest
%that the scaling of $\tau_w$ is independent of the bond concentration $p$ for $p
%< 1$. We believe that these long surviving spins are actually isolated
%spins in the lattice (i.e. spins which have all their bonds removed or
%small groups of spins which have their bonds to the rest of the lattice
%removed). Such spins would only be flipped by the Wolff algorithm if
%they are chosen as the seed spin. And since each spin is equally likely
%to be picked and there are $L^2$ spins, this explains the $L^2$
%limiting factor in the correlation time $\tau_{\mathrm{steps}, w}$.

\subsection{The Swendsen-Wang algorithm}

\begin{figure}
	\includegraphics[width=\columnwidth]{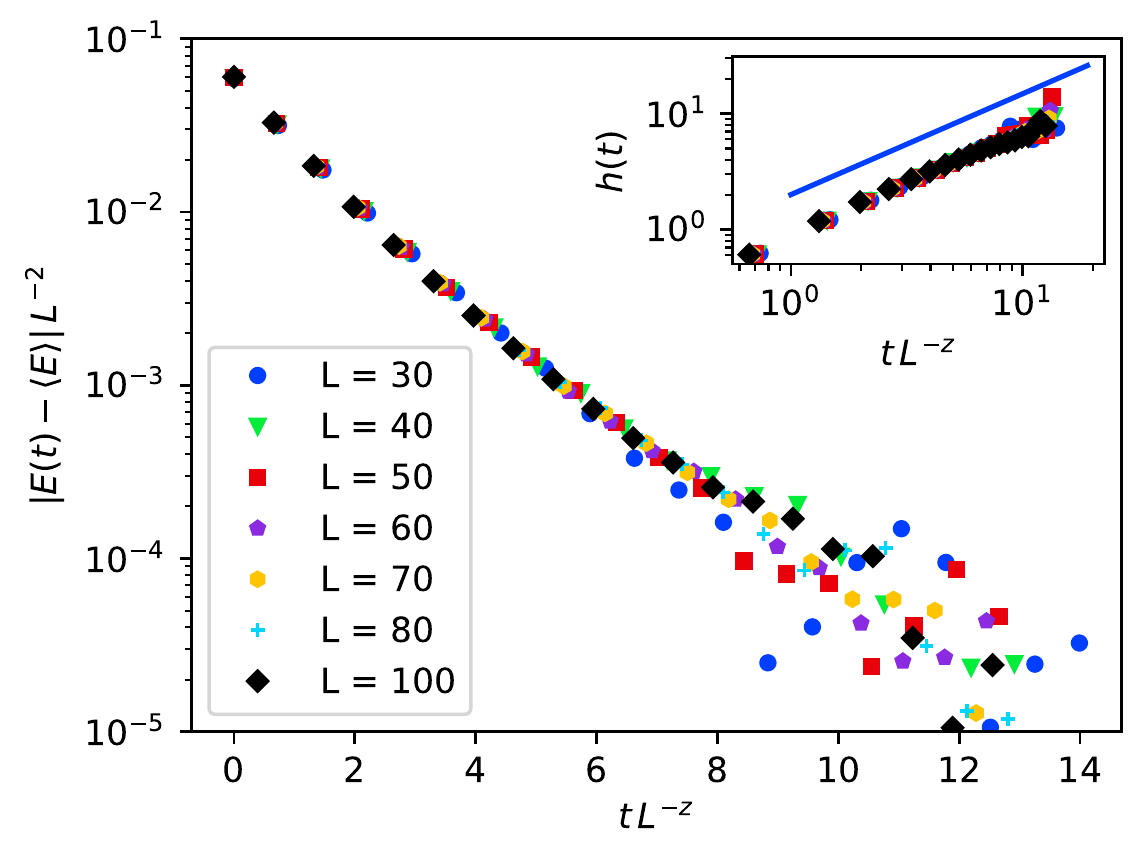}
	\caption{\label{fig:sw_therm} Convergence of the energy $E(t)$ to
	the thermal equilibrium $\langle E \rangle$ during thermalisation
	with the Swendsen-Wang algorithm for different system sizes
	$L$ with $p = 0.6$ at $(\beta J)^{-1} = 0.940$ where $\beta =
	\frac{1}{k_B T}$ and $J$ the coupling constant. For $t = 0$ the
	system starts in a state with all spins pointing up. The vertical
	axis was scaled with $L^2$ and the horizontal axis with $L^{z_{\rm sw}}$
	with $z_{\rm sw} = 0.09(4)$, where $z_{\rm sw}$ was chosen to be the same
	as in figure \ref{fig:sw_energy_othercorr}. Note that this plot is equivalent to figure
	\ref{fig:wolff_therm} but for the Swendsen-Wang algorithm. The
	collapse of the curves suggests that the correlation time for the
	Swendsen-Wang algorithm scales as $L^{z_{\rm sw}}$ with $z_{\rm sw} = 0.09(4)$. Also note the absence
	of a slowly decaying tail, demonstrating that the Swendsen-Wang
	algorithm does not suffer from the same problems that plague
	the Wolff algorithm (see figure \ref{fig:wolff_therm}). The
	inset figure in the top-right shows the same data but plotted
	differently. Here $h(t) = -\log\left(c \left| E(t) - \langle
	E \rangle \right|\right)$ with $c = \left| E(0) - \langle E
	\rangle \right|^{-1}$. The blue curve is a straight line with
	slope $0.87$. Since the data seems to be parallel to this blue
	curve instead of a curve with slope 1, the convergence of the
	energy seems to be stretched exponential.}
\end{figure}

We now turn our attention to the Swendsen-Wang algorithm. By
construction, it visits every spin in the lattice each step, so it should
not suffer from the problems encountered with the Wolff algorithm,
originating from long surviving spins.  Similar to the Wolff
algorithm, we ran simulations for various system sizes $L$ at $p =
0.6$ and $(\beta J)^{-1} = 0.940$, i.e. the setup of the simulations was
exactly the same, only the algorithm used to update the spins was different.
Figure \ref{fig:sw_therm} shows the
analogue of figure \ref{fig:wolff_therm}, but then for Swendsen-Wang. In
addition, it contains an inset figure that shows the same data but plotted
in a different way. At $L = 30$ we ran for 300 Swendsen-Wang steps per
configuration while at $L = 100$ we ran for 100 steps; in between
we tuned the steps to keep the CPU time used roughly constant. In the
main part of the figure we can see that the energy quickly converges to
its thermal equilibrium value and the slowly decaying tail from figure
\ref{fig:wolff_therm} is absent. Moreover, when scaling the vertical
axis with $L^2$ and the horizontal axis with $L^{z_{\rm sw}}$ with $z_{\rm sw} = 0.09(4)$,
the curve collapse suggests that the correlation time $\tau_{\rm sw}$ for
Swendsen-Wang at $p = 0.6$ scales as $L^{z_{\rm sw}}$. The value for $z_{\rm sw}$ used
to scale the horizontal axis was chosen to be the same as the value we determined with
a different method which will be described below. Note that the dynamical
exponent $z_{\rm sw}$ is significantly smaller at $p = 0.6$ than for the regular
2D Ising model ($p = 1$) where $z_{\rm sw} = 0.25(1)$ \cite{BarkemaBook}. This
is the opposite of the super slowing down observed for the
Metropolis algorithm \cite{Zhong_2020}. Finally, in the inset figure
the data for $h(t)$ versus time $t$ is plotted. Here $h(t) = -\log\left(c
\left| E(t) - \langle E \rangle \right|\right)$ with $c = \left| E(0) -
\langle E \rangle \right|^{-1}$. The blue curve is a straight line with
slope $0.87$. Since the data seems to be parallel to this blue curve
instead of a curve with slope $1$, the convergence of the energy seems to
be a stretched exponential.

\begin{figure}
	\includegraphics[width=\columnwidth]{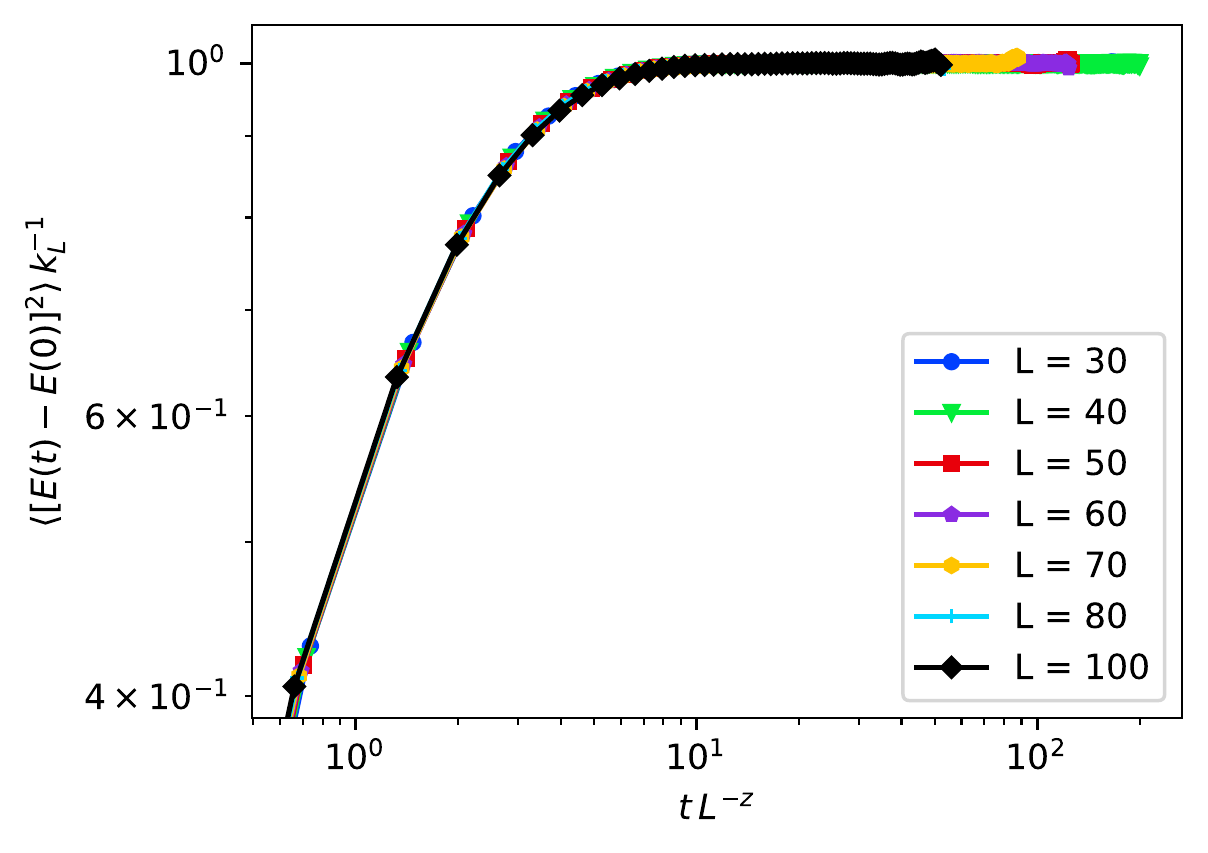}
	\caption{\label{fig:sw_energy_othercorr} Mean-square displacement
	of the energy $\langle [E(t) - E(0)]^2 \rangle$ in thermal
	equilibrium as a function of Swendsen-Wang moves $t$ for
	different system sizes $L$ with $p = 0.6$ at $(\beta J)^{-1}
	= 0.940$ where $\beta = \frac{1}{k_B T}$ and $J$ the coupling
	constant. The vertical axis was scaled with the numerically
	determined limit value of the curves, while the horizontal
	axis was scaled with $L^{z_{\rm sw}}$ with $z_{\rm sw} = 0.09(4)$.
	The dynamical critical exponent for the Swendsen-Wang algorithm
	was determined by tuning the scaling of the horizontal axis until
	a good collapse was found.}
\end{figure}

We have already shown that there is a value for the dynamical exponent $z_{\rm sw}$ that
gives a good collapse of the data in figure \ref{fig:sw_therm}. However,
this plot shows data from simulations out-of-equilibrium so we did not use
this data to determine the correct scaling of the correlation time
(at least not in the form presented in figure \ref{fig:sw_therm}).
Instead, we determined it from equilibrium simulations. For this
we computed the evolution of the mean-square displacement of the
energy $\langle [E(t) - E(0)]^2 \rangle$ from the same data as was used
for figure \ref{fig:sw_therm}. To obtain equilibrium data we discarded
all data before the system was thermalised. For $L = 30$ this concerns all
data before $t = 50$ and for all other system sizes all data before
$t = 20$ (i.e. these times became the new $t = 0$ for determining
$\langle [E(t) - E(0)]^2 \rangle$). The results are shown in figure
\ref{fig:sw_energy_othercorr}. After scaling the vertical axis with
the numerically determined limiting values of the curves, we can collapse
the curves using a horizontal scaling of $L^{z_{\rm sw}}$ with $z_{\rm sw} = 0.09(4)$.
The uncertainty in the dynamical exponent was determined by tuning the scaling of the
axis to determine the range within which the collapse seemed good. The size of this range
was then used as a measure of the uncertainty. This
confirms our earlier numerical estimate of the dynamical critical exponent
for the Swendsen-Wang algorithm at $p = 0.6$.

\section{Summary and Conclusions}

We have shown how the correlation times $\tau_w$ and $\tau_{\rm sw}$ of the Wolff and
Swendsen-Wang cluster algorithms scale as a function of the system
size $L$ when applied to the 2D bond-diluted Ising
model. We demonstrated that the Wolff algorithm suffers from a much
longer correlation time than in the pure Ising model, caused by isolated
(groups of) spins which are infrequently visited by the algorithm. With a simple argument 
we proved that these cause the correlation time to be bounded from below by $L^{z_w}$ where $z_w=\gamma
/ \nu \approx 1.75$ for a bond concentration $p < 1$. Furthermore, we showed numerically that
this lower bound is actually taken for several values of the bond concentration in the region
$0.5 < p < 1$. Moreover, we have shown that the Swendsen-Wang algorithm does not suffer from
the same problem, by construction. It has a much shorter
correlation time, even shorter than in the pure Ising model. Numerically,
we have found that its correlation time scales as $L^{z_{\rm sw}}$ with $z_{\rm sw} = 0.09(4)$
at $p = 0.6$.

% Outlook
We expect that the Wolff algorithm will suffer from the same problems
in the three-dimensional bond-diluted Ising model, albeit to a lesser
degree as more bonds will have to be removed to create isolated spins. In
addition, we think the same will hold for the site-diluted and weakly
diluted (i.e. where you weaken instead of removing the bonds) Ising
models. This could be something to explore in the future.

\bibliography{bibfile}

\end{document}